\documentstyle[12pt]{article}

\newcommand{\bfmath}[1]{\mbox{\boldmath$#1$}}
\newcommand{\vr}{\bfmath r}

\begin{document}

%%%%%%%%%%%%%%%%%%%%%%%%% title page %%%%%%%%%%%%%%%%%%%%%%%

\vskip 20mm

\centerline
{\Large \bf Conservation of Energy   
in Black Holes and in Cosmology}
 
\vskip 20mm 

\centerline{\large Masakatsu Kenmoku
\footnote{E-mail address: kenmoku@phys.nara-wu.ac.jp}}
\centerline{\it Department of Physics, 
Nara Women's University, Nara 630-8506, Japan}

\vskip 3mm 

\centerline{\large Kazuyasu Shigemoto
\footnote{E-mail address: shigemot@tezukayama-u.ac.jp}}
\centerline{\it Department of Physics, 
Tezukayama University, Nara 631-8501, Japan}

\vskip 10mm

\begin{abstract}
We first review the various definition of the total 
energy in the gravitational system. The naive 
definition has some defects, and we review how
to modify the definition of the total energy. 
Then we explicitly demonstrate how to calculate 
the total energy  of the system. 
Our example is the total energy of a black hole 
in the expanding closed de Sitter universe in (2+1) 
dimension. 
In general, we find that the contribution to the 
total energy comes only from the singularity. Then 
we can calculate the total energy by evaluating 
the contribution around the singularity.
\end{abstract}
 
%\pacs{98.80.-k, 98.80.Hw, 97.60.Lf, 95.35.+d}

\vskip 10mm

%%%%%%%%%%%%%%%%%%%%%%% Section 1 %%%%%%%%%%%%%%%%%%%%%%%%
\section{Introduction}

It has a long history how to define the 
total energy in the gravitational theory.
Because of the invariance under the general 
coordinate transformation, it becomes ambiguous 
how to define the time variable. 
Then the Hamiltonian, which is conjugate to time variable, 
becomes ambiguous. Especially there are 
many ways\cite{Einstein,L-L,Moller,Komar}
how to define the energy of the gravitational field. 
The purpose of this paper give the prescription to calculate 
the total energy of the system with local singularity. 
It is known that (i) $E_{\rm total}=({\rm finite})$ 
for the asymptotically flat space-time 
and (ii) $E_{\rm total}=0$ for the closed universe. 
The contribution to the total energy comes only from 
the singularity, so we can calculate the total energy 
by evaluating the contribution around the singularity.

\vspace{0.5\baselineskip}

%%%%%%%%%%%%%%%%%%%%%%%%% Section 2 %%%%%%%%%%%%%%%%%%%%%%%
\setcounter{equation}{0}
\section{Energy of Gravitational Field 
- Review -}
%%%%%%%%%%%%%%%
\subsection{Action and the Einstein's Equation}

The Hilbert-Einstein action is given by \cite{Wheeler}
\begin{eqnarray}
  I=\frac{1}{2 \kappa} 
           \int d^4x \sqrt{-g}\ R 
  +\int d^4x \sqrt{-g} \ {\cal L}_{\rm{matter}}\  , 
\label{e1}
\end{eqnarray}
where $\kappa=8\pi G$.
Taking the variation with respect to $g_{\mu \nu}$, 
we have the Einstein's equation 
$\displaystyle{{\delta I_{\rm HE}}/{\delta g_{\mu \nu}}}=0$,
which gives 
\begin{eqnarray}
  R^{\mu \nu}-\frac{1}{2} R g^{\mu \nu}=\kappa T^{\mu \nu},
\label{e2}
\end{eqnarray}
with the boundary condition
\begin{eqnarray}
  &&\int d^4x \ \partial_{\mu} {\cal B}^{\mu}=0, 
\label{e3}\\
  &&{\cal B}^{\mu}=\sqrt{-g} 
   \left\{ g^{\mu \nu} (\delta \Gamma^{\delta}_{\nu \lambda})
          -g^{\nu \lambda} 
(\delta \Gamma^{\mu}_{\nu \lambda})
   \right\}.
\label{e4}
\end{eqnarray}
Notice that the boundary term ${\cal B^{\mu}}$ contains 
not only the variation of the metric $\delta g_{\mu \nu}$ 
but also the variation of the first derivatives of the metric 
$\delta g_{\mu \nu, \lambda}$ through the affine connection.

%%%%%%%%%%%%%%%
\subsection{Einstein prescription}

In the naive approach, the boundary term ${\cal B^{\mu}}$
contains not only the variation of the metric 
but also the variation of the first derivatives of the metric,
which makes difficult to find the classical solution
compatible with boundary condition. 
There are various approaches to avoid this problem. 

First we review the primitive expression of 
energy conservations derived by Einstein \cite{Einstein} 
in order to avoid the appearance of the variation 
of the derivative of the metric in the boundary.  
Using the identity 
\begin{eqnarray}
 \sqrt{-g} R&=&\sqrt{-g} G  +\partial_\mu {\cal D}^{\mu},
\label{e5} \\
 G&=&g^{\mu \nu} \{ \Gamma^{\rho}_{\mu \nu}
                  \Gamma^{\lambda}_{\rho \lambda}
                - \Gamma^{\lambda}_{\mu \rho}
                  \Gamma^{\rho}_{\nu \lambda} \},
\label{e6}\\
  {\cal D}^{\mu}&=&\sqrt{-g} \{ g^{\mu \nu}
                  \Gamma^{\lambda}_{\nu \lambda}
                - g^{\rho \sigma}
                  \Gamma^{\mu}_{\rho \sigma}\} \ , 
\label{e7}
\end{eqnarray}
and the action is taken in the following form 
\begin{eqnarray}
  I^{*}=\frac{1}{2\kappa} \int d^4x \sqrt{-g} G 
  +\int d^4x \sqrt{-g} {\cal L}_{\rm{matter}} .        
\label{e8}
\end{eqnarray}
After taking the variation of $I^*$ with respect to $g_{\mu \nu}$,   
the same Einstein's equation Eq.(\ref{e2}) is obtained 
with new boundary condition
\begin{eqnarray}
  &&\int d^4x \partial_{\mu} {\cal B}^{* \mu}=0, 
\label{e9}\\
  &&{\cal B}^{* \mu}=-\sqrt{-g} 
   \left\{ \delta (\sqrt{-g} g^{\mu \nu}) 
          \Gamma^{\delta}_{\nu \lambda}
          -\delta( \sqrt{-g} g^{\nu \lambda})
          \Gamma^{\mu}_{\nu \lambda} 
   \right\}   .
\label{e10}
\end{eqnarray}
In this expression, the boundary term ${\cal B}^{* \mu}$
contains only the variation of the metric as expected. 
Putting $\displaystyle{
{\cal L}^{*}_{\rm{G}}=  \sqrt{-g} G /(2\kappa)}$,
the energy-momentum tensor of the gravitational field 
is given by 
\begin{eqnarray}
u^{\mu}_{\nu}=\frac{\partial {\cal L}^{*}_{\rm{G}}}
                {\partial g^{\rho \sigma}_{,\mu} }
   g^{\rho \sigma}_{,\mu} -\delta^{\mu}_{\nu} 
   {\cal L}^{*}_{\rm{G}}.
\label{e11}
\end{eqnarray}
Then the conservation law of energy-momentum is expressed 
as 
\begin{eqnarray}
\partial_{\mu} (u^{\mu}_{\nu}+T^{\mu}_{\nu})=0,
\label{e12}   
\end{eqnarray}
where $T^{\mu}_{\nu}$ denotes the energy-momentum tensor 
density of matter. 
However the separation of the dynamical metric  
from the auxiliary metric is not clear in this method. 
  
%%%%%%%%%%%%%%%%%%%%%%%%%
\subsection{ADM canonical formalism}

In order to clarify the physical meaning of the metrics,  
we often use the ADM formalism \cite{ADM} where
we decompose the 4-dimensional space-time metric into 
(3+1) dimensional space and time metric.
We consider the Cauchy surface $\Sigma_t$ parametrized by 
a global function of time $t$. Let $n^a$ be the unit normal 
vector to the hypersurface $\Sigma_t$.
The metrics are decomposed into the form
\begin{eqnarray}
ds^2&=&-N^2 dt^2+h_{a b}(dx^{a}+N^{a})(dx^{b}+N^{b}),
\label{e13}\\
g_{a b}&=&\left(\begin{array}{cc}
              -N^2+N^a N_b , N_a \\
              \quad   N_a \quad \quad ,h_{a b} \\
              \end{array} \right) , 
\label{e14}
\end{eqnarray}
where $N$ is the lapse function, $N^a$ is the shift operator 
and $h_{ab}$ is the spatial metric. 
In this expression, the determinant of the metrics is 
$g= Nh$.

First we proceed the Lagrangian formulation under the 
ADM (3+1) decomposition. 
Defining the extrinsic curvature of $\Sigma_{t}$ 
\begin{eqnarray}
 K_{a b}=h^{c}_{a} \nabla_{c} n_b
   \quad (= N \Gamma^{0}_{a b}) 
    \label{e15}
\end{eqnarray}
and its mean value $K=K_{a}^{a}$, 
the gravitational action is taken in the form
\begin{eqnarray}
  I_{\rm G}&=&\frac{1}{2 \kappa} 
(\int d^4x  \sqrt{-g} R +\int d^3x \sqrt{h} 2K ) ,  
\label{e16}\\
 \ &=&\frac{1}{2 \kappa} 
   \int d^4x N \sqrt{h}(K_{a b}K^{a b}-K^2+^{(3)}R) .  
\label{e17}
\end{eqnarray}
The notation $(3)$ in the left upper indices  
denotes the corresponding three dimensional quantities. 
Taking the variation of $I_{\rm G}$ we have the Einstein's equation 
with the boundary condition, which contains only the variation of the 
first time derivatives of metrics. 
The surface term $\int dx^4 \partial_t K /\kappa$ plays the role to make  
the action well defined in the Euclidean section  \cite{GH}. 
%%%%%%%%%%%%%%%%%%%%%%%%%%%%%
%\subsection{Hamiltonian formulation} 

Next we proceed to the Hamiltonian formulation. 
The canonical momentum 
and the Hamiltonian density are expressed as 
\begin{eqnarray}
 \pi^{a b}&=&\frac{\partial {\cal L}_{\rm G}}
               {\partial \dot{h}_{a b}}
   =\sqrt{h}(K^{a b}-K h^{a b}),   
\label{e18}\\
{\cal H_{\rm G}}&=&\pi^{a b} \dot{h}_{a b}-{\cal L}_{\rm G} \nonumber\\
 &=&N H+N^a P_a+2 D_a (h^{-1/2} N_b \pi^{a b}). 
\label{e19}
\end{eqnarray}
In the above expression, 
the last term contribute only to an unimportant boundary term 
and will be dropped in the following. 
The Hamiltonian function and momentum function are given by 
\begin{eqnarray}
H&=&\frac{1}{\sqrt{h}}(\pi^{a b} \pi_{a b} 
  -\frac{1}{2}\pi^2)-\sqrt{h}^{(3)}R  ,
\label{e20}\\
P_a&=&-2 \sqrt{h}( D_a h^{-1/2} \pi^{a b}) .
\label{e21}
\end{eqnarray}
Because the time derivatives of $N$ and $N^a$ do not appear
in the Lagrangian, they are auxiliary variables. Then
we must put the Hamiltonian and the momentum constraints as 
\begin{eqnarray}
H \approx 0 ,
     \label{e22}\\
P_a \approx 0 .
     \label{e23}
\end{eqnarray}
The dynamical equations, which are derived by the canonical 
formalism, are given by
\begin{eqnarray}
   \dot{h}_{a b} &&=\ \ \frac{\delta H_{\rm G}}{\delta \pi^{a b}} 
   \equiv \ \ A_{a b},   \label{e24}\\
   \dot{\pi}^{a b} &&=-\frac{\delta H_{\rm G}}{\delta h_{a b}} 
    \equiv -B_{a b} ,  \label{e25}
\end{eqnarray}
where 
\begin{eqnarray}
        H_{\rm G}=\int d^3x ( NH+N^a P_a ).   \label{e26}
\end{eqnarray}
While, if we directly take the variation of this ADM 
Hamiltonian  
\begin{eqnarray}
  \delta H_{\rm G}&&=\int d^3x ( A_{a b} \delta \pi^{a b}
             + B^{a b} \delta h_{a b} ) - \delta C,    
     \label{e27}
\end{eqnarray}
we obtain an extra surface contribution $\delta C$. 
So the dynamical equation derived by the canonical formalism 
and the dynamical equation derived by the Lagrangian formalism 
give the different boundary condition.
 
%%%%%%%%%%%%%%%%%%%
%\subsection{Regge-Teitelboim interpretation }

In order to avoid the additional surface term
$\delta C$, Regge-Teitelboim \cite{Regge} modify the total 
Hamiltonian 
in such a way as the dynamical equation, derived by the 
canonical formalism, is equivalent to the equation 
derived by the variation of the total Hamiltonian. 
The explicit form of $\delta C$ is  
\begin{eqnarray}
     \delta C =
    && \oint d^2S_l G^{i j k l} 
       (N \delta h_{ij, k}-N_{,k} \delta h_{ij})
       \nonumber\\
    && + \oint d^2S_l \{ 2 N_{k} \delta \pi^{kl} 
          +(2N^k \pi^{j l}-N^l \pi^{j k}) \delta h_{jk}\} ,
     \label{e28}
\end{eqnarray}
where
\begin{eqnarray}
   G^{i j k l}=\frac{1}{2}h^{1/2} 
       (h^{ik} h^{jl} +h^{il} h^{jk}-2 h^{ij} h^{kl}).
     \label{e29}
\end{eqnarray}
If this surface term would vanish, $H_{\rm G}$ would be the correct 
Hamiltonian. However for the stationary asymptotically flat space-time  
$(h_{ij} - \delta_{ij}\sim 1/r )$, 
the first surface term is different from zero,  
 that is 
\begin{eqnarray}
\oint d^2S_l G^{i j k l} 
       (N \delta h_{ij, k}-N_{,k} \delta h_{ij})
     \sim   \delta \oint d^2S_l (h_{il,i}-h_{ii,l}).  
       \label{e30}
\end{eqnarray}
A new gravitational Hamiltonian is defined by \cite{Wald} 
\begin{eqnarray} 
H_{\rm G}' =H_{\rm G}+E[h_{ij}], 
      \label{e31}
\end{eqnarray} 
where additional contribution comes from the surface term by   
\begin{eqnarray}
E[h_{ij}] =\oint d^2 S_{l}  G^{i j k l} h_{ij,k}.
  \label{e32}
\end{eqnarray}
This is called ADM energy for the stationary 
asymptotically flat space-time. 
We get for the variation in this new gravitational Hamiltonian
\begin{eqnarray}
  \delta H_{\rm G}' &&=\int d^3x ( A_{a b} \delta \pi^{a b}
             + B^{a b} \delta h_{a b} ),     
     \label{e33}
\end{eqnarray}
and the correct equations of motion Eq.(\ref {e24})-Eq.(\ref {e25}) 
are recovered. 

%%%%%%%%%%%%%%%%%%%%%
%\subsection{Value of gravitational energy}

The value of the energy becomes in the following.
First, one has $H_{\rm G}=0$ because of the constraint 
equations Eq.(\ref{e20})-Eq.(\ref{e21}).
For the surface contribution one has $E[h_{i j}]=0$ for the closed universe,
because there is no boundary, and 
$E[h_{i j}]=({\rm finite})$ for the asymptotically flat space-time.
For example, for the Schwarzschild metric, 
\begin{eqnarray}
ds^2 \sim -(1-\frac{2Gm}{r})dt^2 
          + (\delta_{ij}+2Gm \frac{x^i x^j}{r^3})dx^i dx^j , 
\nonumber 
\end{eqnarray}
one has well known result $E[h_{ij}]=m$ \cite{Komar,DeWitt-Kuchar}.

%%%%%%%%%%%%%%%%%%%%%%%
\subsection{Comment on the surface term}

The surface terms in Eq.(\ref{e16}) and Eq.(\ref{e32}) 
are important for 
the Lagragian formalism and the Hamiltonian formalism. 
Here we comment these surface terms are related to 
the surface term appeared in the identity Eq.(\ref{e5})-Eq.(\ref{e7}). 
\\
A. \  The surface term, which is the mean value of the 
scalar curvature  in Eq.(\ref{e16}),  
plays a similar role to the time component of the surface term 
Eq.(\ref{e7}) in Einstein's prescription in order to avoid 
the time derivative contribution in the surface term; 
\begin{eqnarray}
h^{1/2} 2 K \sim {\cal D}^{t} \nonumber. 
\end{eqnarray}
B. \  
Using the three dimensional version of the identity Eq.(\ref{e5})
-Eq.(\ref{e7})   
\begin{eqnarray}
 \sqrt{h}^{(3)}R&=&\sqrt{h}^{(3)} G  
+\partial_l{\cal^{\rm (3)} D}^{l},
\label{e33} \\
 {\cal ^{\rm (3)}D}^{l}&\equiv&\sqrt{h} 
\{- h^{li \ (3)}\Gamma ^{j}_{ij}
               + h^{ij\ (3)}\Gamma^{l}_{ij} \} \nonumber \\
&=& ^{(3)} G^{ijkl}h_{ij,k}, 
\label{e35}
\end{eqnarray}
where 
\begin{eqnarray}
^{(3)}G^{ijkl}\equiv \frac{1}{2}h^{1/2}\{h^{ik}h^{jl} +h^{il}h^{jk} 
-2h^{ij}h^{kl}\} , 
\label{e36}
\end{eqnarray}
the additional energy Eq.(\ref{e32}) contributed from 
the surface term is simply expressed as
\begin{eqnarray}
E[h_{ij}] = \frac{1}{2\kappa}
  \int d^3 x \partial_l {\cal ^{\rm (3)}D}^{l} . 
\label{e37}
\end{eqnarray}

From the above observation  
the contribution to the total energy is interpreted 
in the following way. 
The term $\sqrt{-g} R$ includes the singular terms 
coming from second derivative terms, and 
$\sqrt{-g} G$ contains the less singular terms 
including only first derivative terms and the term 
$\partial_\mu {\cal D}^{\mu}$ picks up the singular part 
as surface terms.

Next we study the cases of neither closed space nor 
asymptotically flat space-time. 

\vspace{0.5\baselineskip} 

%%%%%%%%%%%%%%%%%%%%%%%%% Section 3 %%%%%%%%%%%%%%%%%%%%%%%
\section{Energy of System with Conical Singularity}
%%%%%%%%%%%%%%%
\subsection{(2+1) dimensional conical singularity}

We consider the system of the point particle 
with mass $m_0$ \cite{DJH}.
Static spherically symmetric metric is given by
\begin{eqnarray}
ds^2=-dt^2+{\rm e}^{\sigma(\vr)} (dr^2+r^2 d\phi^2).
\label{e38}
\end{eqnarray}
Einstein equation becomes  
\begin{eqnarray}
 \triangle \sigma(\vr)=m_0 \delta(\vr).
\nonumber
\end{eqnarray}
Then the solution is given as  
\begin{eqnarray}
ds^2=-dt^2+r^{-8 G m_0 }(dr^2+r^2 d\phi^2) ,
\label{e39}
\end{eqnarray}
where the ranges of variables are 
$0 \le r < \infty$ and $0 \le \phi \le 2 \pi$.
After making the change of variables 
\begin{eqnarray}
\bar{r}={r^{1-4 G m_0 }}/(1-4 G m_0 ), \quad
\bar{\phi}=(1-4 G m_0 ) \phi . \label{e40}
\end{eqnarray}
we get the flat metric   
\begin{eqnarray}
ds^2=-dt^2+d \bar{r}^2+\bar{r}^2 d\phi^2 , \label{e41}\\
 0 \le  \bar{r} <\infty , \quad   
0 \le \bar{\phi} \le 2 \pi -\Delta , \label{e42}
\end{eqnarray}
where the deficit angle $\Delta=8 \pi G m_0$ appears 
in the range of variable $\phi$.
This metric is not asymptotically flat and the total 
energy formula Eq.(\ref{e31}) cannot apply. 
Here we define the total energy of this system as  
\begin{eqnarray}
E_{\rm total}= -\frac{1}{2 \kappa} \int d^3 x H 
   \label{e43}, 
\end{eqnarray}
where $H$ is defined in Eq.(\ref{e20}). 
Note that the minus sign, three dimensional integration and 
lack of the lapse function $N$ in $E_{\rm total}$ of 
Eq.(\ref{e43}).   
Using the explicit expression for $H$in Eq.(\ref{e20}), 
the correct value of total energy is obtained \cite{DJH} 
\begin{eqnarray}
E_{\rm total}=\frac{1}{2 \kappa} \int d^2 x \sqrt{h}
    ^{(2)}R
     =-\frac{1}{2 \kappa} \oint dS_l \nabla^{l} \sigma(x)
   =m_0 .    \label{e44}
\end{eqnarray}
Next we explore the gravitational energy for the 
case with a singularity in the closed universe.  
%%%%%%%%%%%%%%%
\subsection{Closed and expanding de Sitter universe with 
 conical singularity in (2+1) dimension}

We consider the system of closed and expanding de Sitter universe 
with conical singularity. 
The classical solution is given by \cite{KMU}
\begin{eqnarray}
ds^2=-dt^2+a(t)^2 \left( \frac{d \bar{r}^2 }{ 1-\bar{r}^2}
+(1-8 G m) \bar{r}^2 d\phi^2 \right) .
\label{e45}
\end{eqnarray}
Here the scale factor of the universe $a(t)$ is determined by the 
equation 
$ {\dot{a}^2}/{a^2}=-{1}/{a^2}+ \lambda$ , 
and the solution is given by 
$ a(t)= \cosh( \sqrt{\lambda} t)/ \sqrt{\lambda} $. 
After making the change of variable 
$\bar{\phi}=(1-b)\phi$, 
the metrics in Eq.(\ref{e45}) shows the appearance of the 
deficit angle $\Delta = 2\pi b \sim 8\pi G m $, 
where the parameter $b=1- \sqrt{1-8 G m}$ is introduced. 
In order to evaluate the total energy of the system 
we make the change of variables 
\begin{eqnarray}
 \bar{r}= {r^{1-b}}/(1-b) , \nonumber 
\end{eqnarray}
and we get the metric in the form   
\begin{eqnarray}
ds^2 =-dt^2+a(t)^2 r^{-2b}
  \left(  \frac{d r^2 }{ 1-r^{2-2b}/(1-b)^2}
    +r^2 d\phi^2 \right).
 \label{e46}
\end{eqnarray}
The total gravitational energy is defined 
as the same form in Eq.(\ref{e43}). 
Almost all region in the integration vanish due to the 
constraint Eq.(\ref{e22}) and only contribution comes from 
the singular part around $r=0$. Therefore the integration 
is performed small region $\epsilon$ around $r=0$ 
and the total gravitational energy is obtained as 
\begin{eqnarray}
E_{\rm total}
&=&-\frac{1}{2\kappa} \int 
\frac{1}{\sqrt{h}}(\pi^{a b} \pi_{a b}
   -\frac{1}{2} \pi^2) -{\sqrt{h}}^{(2)}( ^{(2)}R-2 \lambda)
\nonumber \\ 
&=&-\frac{1}{\kappa} \int^{\epsilon}_{0} dr
 \int^{2 \pi}_{0} d\phi 
   \frac{ \partial}{\partial r} 
   \left(  \sqrt{h} h^{rr}  \frac{\partial}{\partial r}
    \sqrt{h_{\phi \phi}} \right) \nonumber \\
   &=& 2m ,  
 \label{e47}
\end{eqnarray}
where $\lambda$ is the cosmological constant. 
Note that the value of the total energy for the closed 
expanding de Sitter solution with conical singularity 
in Eq.(\ref{e47})
is twice for the simple conical singular solution  
in Eq.(\ref{e44}). 

The gravitational energy in asymptotically de Sitter spaces 
is examined by Abbott and Deser \cite{AD82}. 
The three dimensional cosmological gravity is studied 
by Deser and Jackiw and found 
static many-body solutions \cite{DJ84}.   
Mass in static asymptotically de Sitter spaces was 
studied by V. Balasubramanian et al. \cite{BBM02}.  
\vspace{0.5\baselineskip} 

%%%%%%%%%%%%%%%%%%%%%%%%%%%% Section 4 %%%%%%%%%%%%%%%%%%%%%%
\setcounter{equation}{0}
\section{Summary and Discussion}

We first review the derivation of the total gravitational 
energy,
i)Einstein prescription, 
ii) ADM canonical formalism using 
 Regge-Teitelboim derivation.

In section 3, we explicitly calculate the total 
gravitational energy for neither closed space nor 
stationary asymptotically flat space-time. 
Examples 
are the following (2+1) dimensional systems, 
i) system with conical singularity, 
ii) closed and expanding de Sitter universe 
with conical singularity. 

In general, we find 
that the contribution to the total energy comes 
only from the singularity. Then we can calculate 
the total energy by evaluating the contribution 
around the singularity.
We will show the complete proof for our statement near future. 

%%%%%%%%%%%%%%%%%%%%%%%%% references %%%%%%%%%%%%%%%%%%%

\end{document}